\documentclass{sig-alternate}

\usepackage{graphicx,latexsym,epsf,epsfig,epstopdf}
\usepackage{amssymb,amsmath}
\usepackage{alltt}
\usepackage{enumitem}
\usepackage{amsfonts}
\usepackage{pifont}
\usepackage{algorithm, algorithmic}
\usepackage{float}
\usepackage{subfigure}
\usepackage[hyphens]{url}

\usepackage{color}
\usepackage{xspace}

\newcommand{\ours}{\texttt{Anima}\xspace}

\newcommand{\cmark}{\ding{51}}%
\newcommand{\xmark}{\ding{55}}%

\def\sharedaffiliation{
\end{tabular}
\begin{tabular}{c}}

\def\stopproof{\square}
\def\square{\vbox{\hrule height.2pt\hbox{\vrule width.2pt height5pt \kern5pt
\vrule width.2pt} \hrule height.2pt}}

\begin{document}

\title{Anima: Adaptive Personalized Software Keyboard}

\numberofauthors{4}
\author{
\alignauthor
Panos Sakkos 2\\
\email{panoss@microsoft.com}\thanks{Work done while at Unversity of Athens}
\alignauthor
Dimitrios Kotsakos 1\\
\email{dimkots@di.uoa.gr}
\alignauthor 
Ioannis Katakis 1\\
\email{katak@di.uoa.gr}
\and
\alignauthor 
\email{}
\alignauthor
Dimitrios Gunopulos 1\\
\email{dg@di.uoa.gr}
\alignauthor 
\email{}
\sharedaffiliation[1]
      \affaddr{Dept. of Informatics and Telecommunications} \\
      \affaddr{University of Athens, Ilissia GR15784, Greece.}
\sharedaffiliation[2]
      \affaddr{Microsoft Development Center Norway} \\
      \affaddr{Torggata 2-4-6, PB 043, Oslo 0104, Norway.}
}

\maketitle

\begin{abstract}
We present a Software Keyboard for smart touchscreen devices that learns its owner's unique dictionary in order to produce personalized typing predictions. The learning process is accelerated by analysing user's past typed communication. Moreover, personal temporal user behaviour is captured and exploited in the prediction engine. Computational and storage issues are addressed by dynamically forgetting words that the user no longer types. A prototype implementation is available at Google Play Store.
\end{abstract}
\category{H.2.8}{Database Applications}{Data mining}
\category{C.5.3}{Computer System Implementation}{Personal Computers}

\terms{Algorithms}

\keywords{software keyboards, text predictions} 

\section{Introduction}\label{sec:intro}
Over the last few years smartphones and tablets, have achieved mass adoption. Currently, smart devices possess impressive computational and sensing capabilities. However, little improvement has been achieved with respect to their (software) keyboards (SKs). 

Contemporary SKs are more difficult to use, due to lack of typing feedback, large number of keys and small screen size. On the other hand, the need for written communication is increasing. Popular everyday applications include chat messaging, e-mail exchanging, micro-blogging, participation in social networks and even editing documents. 

With the rise of multitouch capacitive touch-screens, various research approaches \cite{yin2013making,
bi2013octopus, rodrigues2013improving} and corporate efforts \cite{apple:ios} tried to improve software keyboards. However, none of them managed to be successfully adopted yet. Existing commercial SKs, are based on generic dictionaries \cite{apple:ios, android:dic} adapting to personal typing behavior by learning new words and phrases that the user commonly uses. Generic predefined dictionaries maintain a large amount of unnecessary information and fail to capture the user language. For example, it is common for non-English speakers to type their messages in their mother tongue using Latin characters. These dialects use words that can not be found in common dictionaries, since they differ from user to user and there is no common grammar. As a consequence, `auto-complete' solutions have  disappointed users even leading them into creating memes discussing these issues \cite{damn:2014}. 

We introduce \ours, an SK that captures its owner's unique dictionary in order to produce personalized typing predictions. \ours is characterized by the following \emph{novel} features:

\begin{enumerate}
	\item There is no requirement for a predefined dictionary
	\item The learning process is accelerated by considering the user's past written communication
	\item A low memory footprint is achieved, by forgetting words that the user does not use any more, and finally
	\item \ours learns user's daily patterns and produces time-aware predictions
\end{enumerate}

\section{System Overview}

\noindent\textbf{Architecture.} According to Human Computer Interaction (HCI) research, \emph{word} suggestion engines (`auto-complete') decrease the user's typing speed  without reducing error rates \cite{rodrigues2013improving}. Hence, we focus on predicting the \emph{next character} that the user will most probably type. The architecture of \ours decouples the User Interface (UI) from the Prediction Engine (PE). In this way, we provide the opportunity to the HCI research community to develop new UI approaches that will utilize the existing PE. Hence, we defined a simple contrac that the UI must follow: 
\begin{enumerate}
  \item Inform PE about the typed character 
  \item Request predictions for the next character 
  \item Visually exploit the predicted character set that PE returns 
  \item Send feedback to PE when a prediction is not accurate
\end{enumerate}

\begin{figure}[ht]
  \centering
  \includegraphics[width=2.6in,natwidth=937,natheight=927]{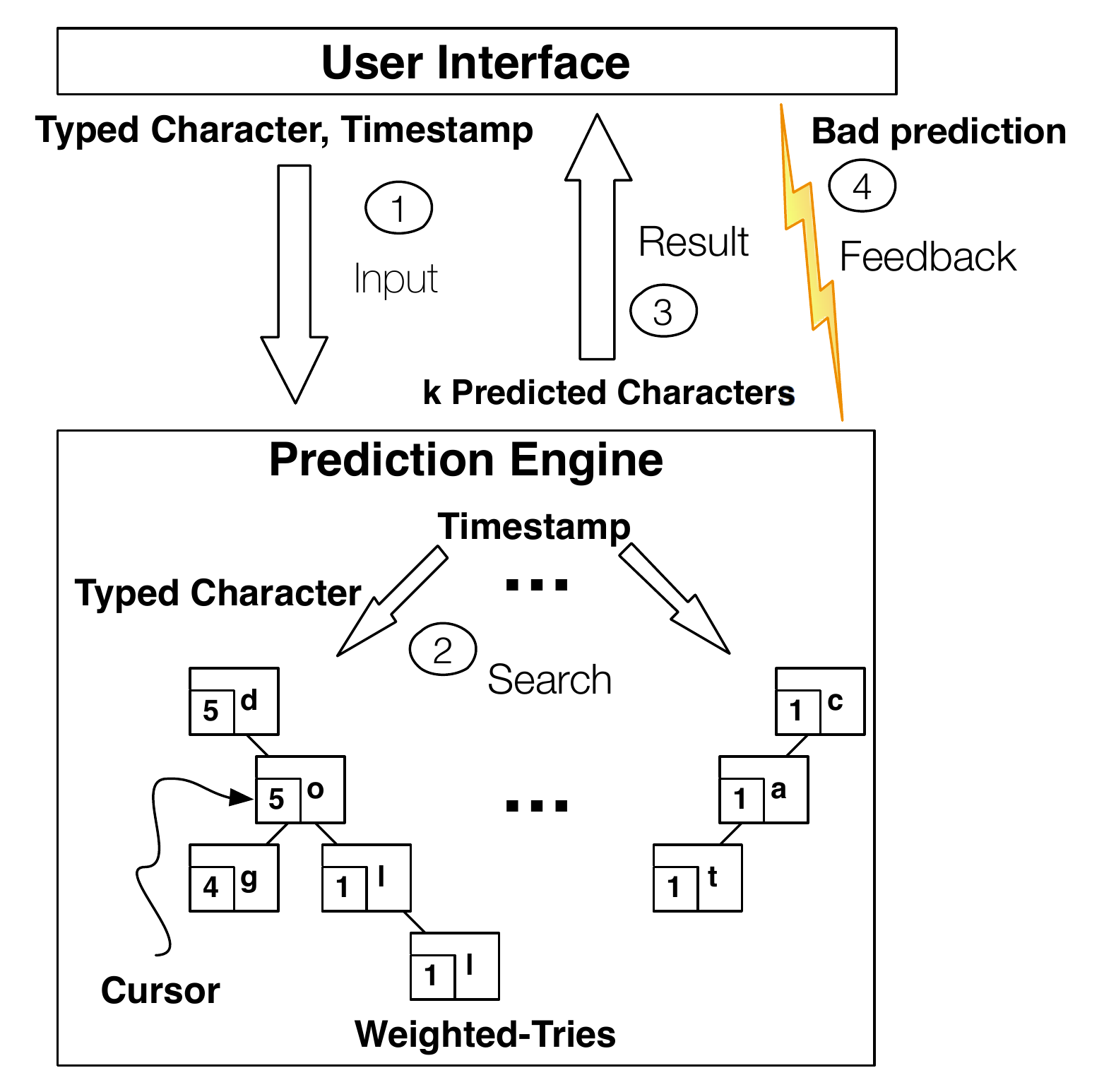}
  \caption{\ours overview}
  \label{fig:sys}
\end{figure}

The system overview of \ours is depicted in Figure \ref{fig:sys}.\\

\begin{figure}
  \centering
  \fbox{\includegraphics[scale=0.21,natwidth=480,natheight=800]{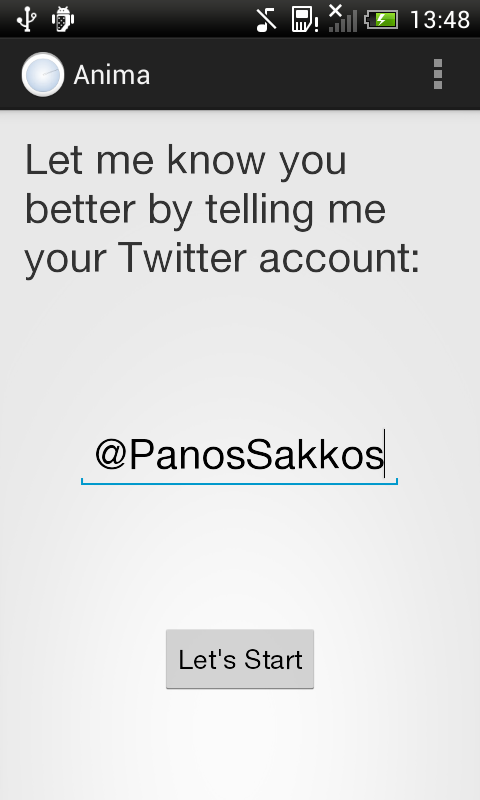}}
  \caption{\ours \textit{Twitter Accelerator}}
  \label{fig:anima}
\end{figure}

\noindent\textbf{User Interface.} \ours's UI, is inspired by recent HCI research \cite{gunawardana2010usability} \emph{shrinking} the keyboard buttons that are not included in the PE's prediction set. As a result, it visually aids the user to type the desired characters (see Figure \ref{fig:anima}). In addition, shrinking buttons result to larger empty spaces between keys preventing mistypes. In order send feedback for a bad prediction the user can swipe diagonally. Finally, the user can hide the keyboard by swiping down.

\begin{figure*}
  \centering
  \subfigure[\textit{""}]{\fbox{\includegraphics[scale=0.1,natwidth=1080,natheight=589]{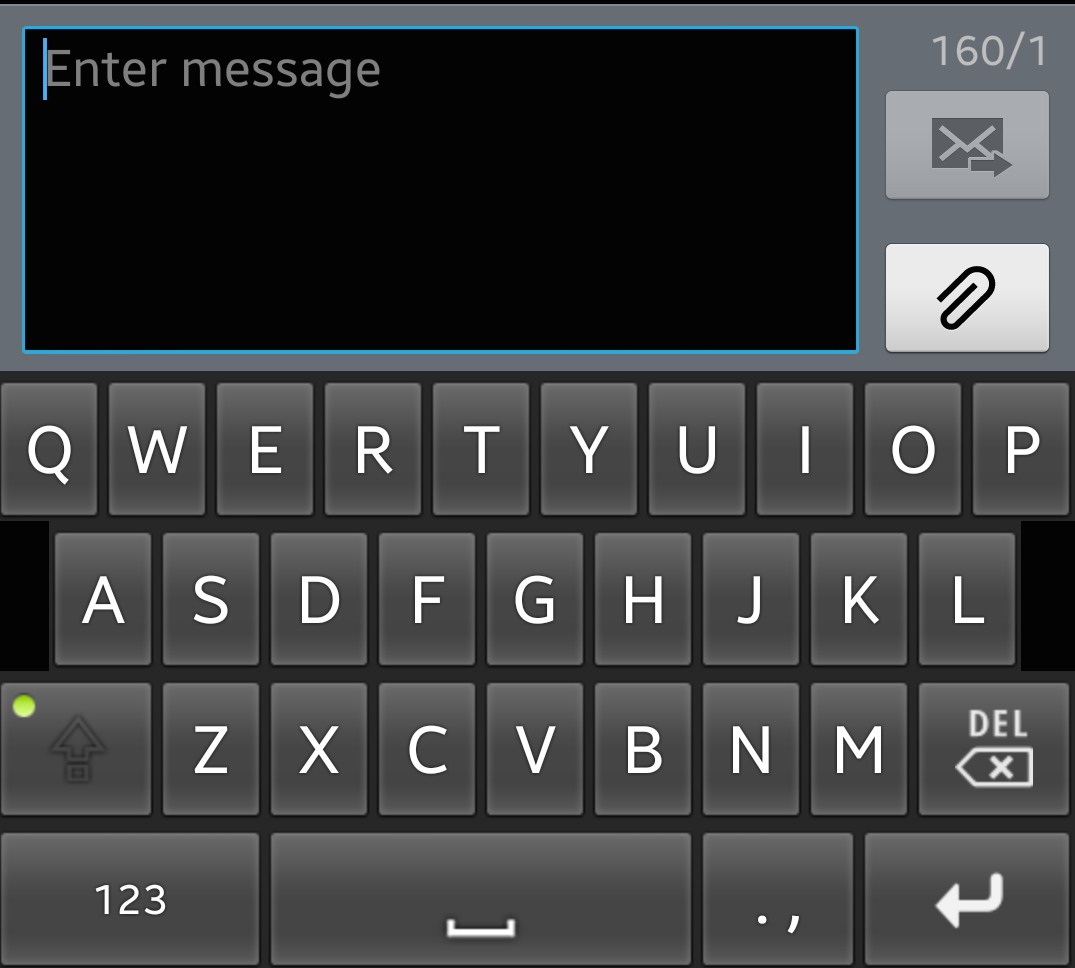}}}
  \subfigure[\textit{"D"}]{\fbox{\includegraphics[scale=0.1,natwidth=1080,natheight=597]{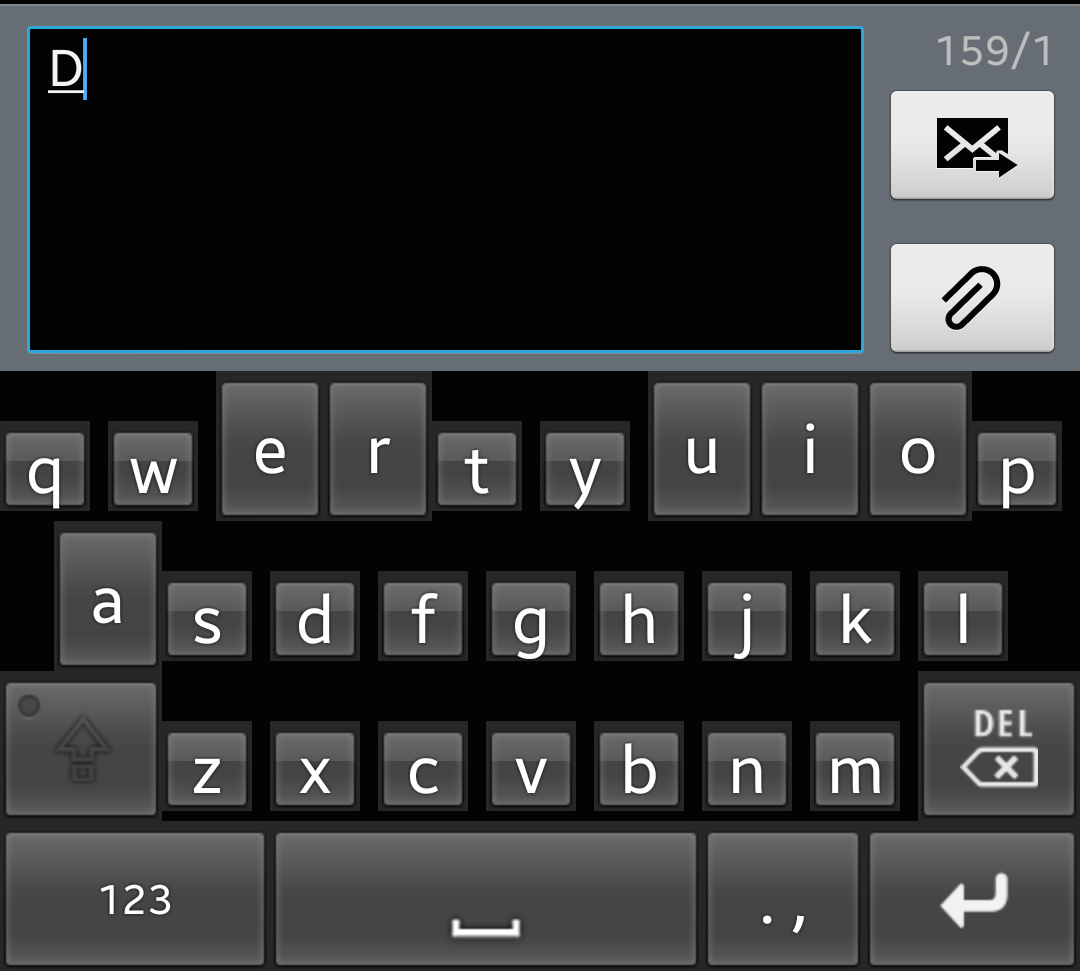}}}
  \subfigure[\textit{"Do"}]{\fbox{\includegraphics[scale=0.1,natwidth=1080,natheight=597]{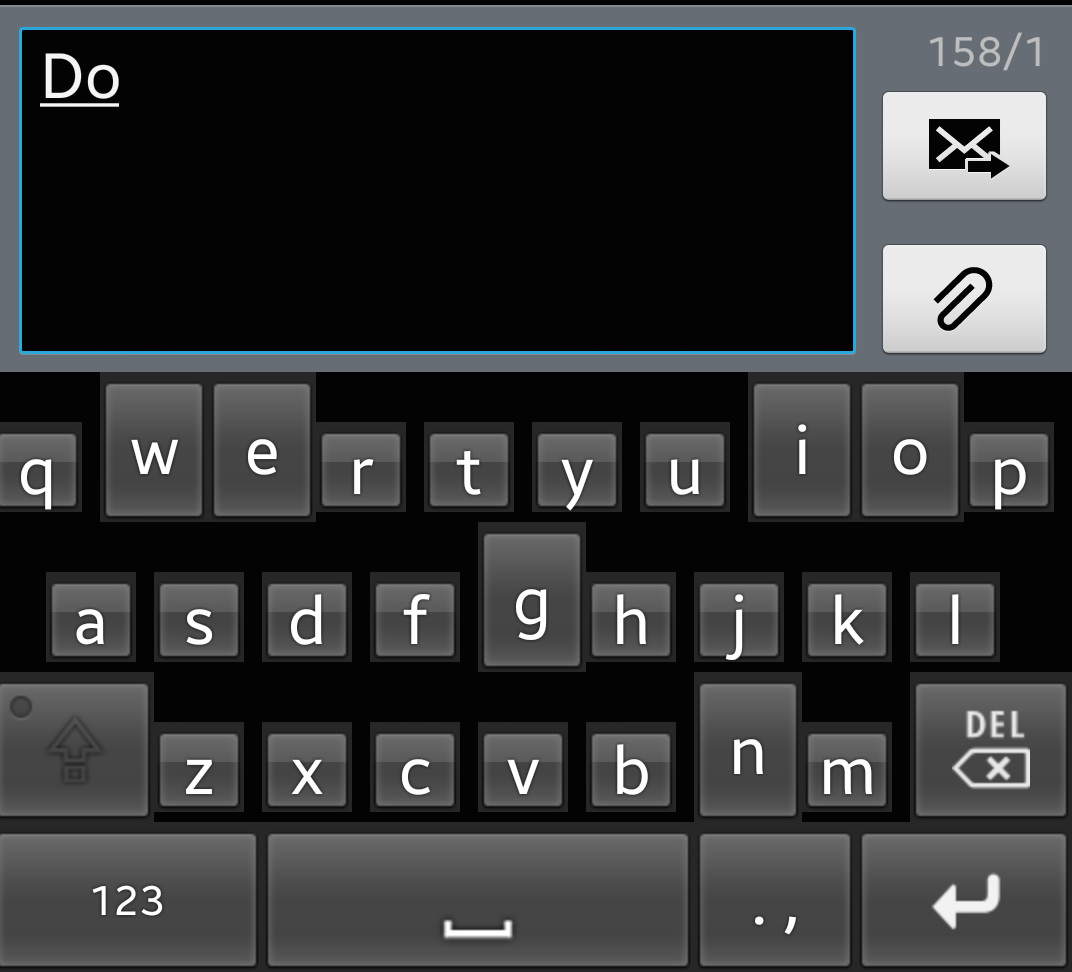}}}
  \subfigure[\textit{"Dog"}]{\fbox{\includegraphics[scale=0.1,natwidth=1079,natheight=600]{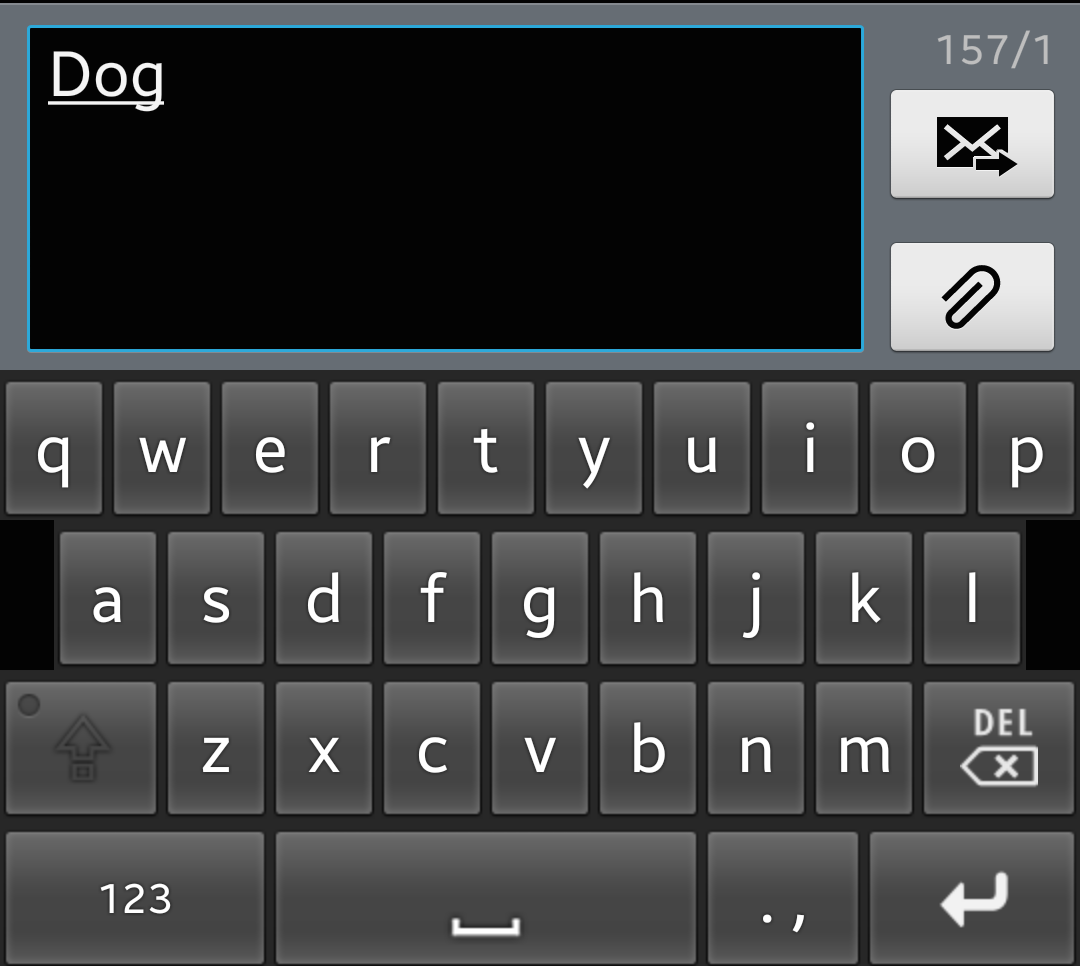}}}
\caption{\ours UI interaction while user is typing the word \textit{Dog}}
\label{fig:animaui}
\end{figure*}

\section{Prediction Engine}
\label{sec:pe}
The Prediction Engine is the module that predicts the next character that the user will type. Its main structure is based on a \emph{Trie} data structure \cite{willard1984new}, with weights in each node. 
Let's assume a stream of user key strokes at each time-step $t$: $\{a_1,\ldots ,a_t, \ldots\}$, where $a_t \in K$ and $K$ is the set of \emph{all} possible keys in a keyboard. A corresponding stream of prediction-\emph{sets} is produced by the prediction engine $\{G_1, \ldots, G_t, \ldots \}$, where $G_t=\{g_{(t,1)},\ldots,g_{(t,n_t)}\}$, with $g \in K$ and $n$ the size of the prediction set. Note that $t$ in $n_t$ implies that the size of the prediction set is varying (see next paragraphs).
  
\textbf{Learning and Predicting.} Each time the user types a character $a_t$, PE traverses the Trie from the root, by descending one level every time a character is typed. The weights in each node, represent how many times the user has traversed this path. The weighted Trie ($WT$) tracks the traversed paths and adds words that the user types for the first time. A cursor (\texttt{c}) points to the current level of the Trie. Every time the user types a word separator, $c$ is redirected to the root of $WT$. In order to produce predictions for the next character $G_{t+1}$, given the sequence of characters that have been typed after the last word separator (word $W$), \ours computes the conditional probabilities $P(k \mid W)$ for every $k\in K$, based on the weights of the nodes at $c$. Then, the top-$n_t$ characters according to $P$ are returned. The time complexity of computing all probabilities of the next character is constant, $O(1)$. This results from the following:

\begin{itemize}
  \item The limited number of probability calculations (i.e. size of $K$) and 
  \item The fact that these computations consider information that is available within the node of $WT$ (instant access)
\end{itemize}

Constant time complexity is crucial for such applications, since any propagated delay to the UI level, would annoy the user and hurt the usability of the keyboard.

\noindent \textbf{Confidence and Diffidence.} An upper bound $n_t$ is defined, representing the maximum number of predicted characters at time $t$. This upper bound is dynamic and is refined automatically and continuously. It is associated with the confidence of the PE. The time that $n_t$ is adjusted, is defined by two parameters, \texttt{conf} (confidence) and \texttt{diff} (diffidence) associated with bad prediction feedback from the UI. After \texttt{conf} consecutive successful predictions, $n_t$ is decreased by one, resulting to a more aggressive (confident) behaviour shrinking $k-n_t$ keys. On the other hand, when PE gets \texttt{diff} continuous negative feedbacks, $n_t$ is increased by one, in order to shrink a smaller set of characters.

\noindent \textbf{Learning Accelerators.} In order to accelerate the learning process, \emph{Accelerators} can be added to the PE. \emph{Accelerators} are messages written by the user in the past. Currently one \emph{Accelerator} has been implemented, shown in Figure \ref{fig:anima}, which crawls the user's Twitter account and feeds his/her messages to the engine, in order to train it.

\noindent \textbf{Pruning.} Contemporary SKs come with a pre-installed dictionary and learn words that the user is typing and are not included in the dictionary. Google's Android SK contains a dictionary of 160,722 words for the English UK language \cite{android:dic} and its footprint continues to increase as the user types new words (see Figure \ref{fig:pre_words}b). In order to address the increasing volume of the dictionary, \ours prunes $WT$, based on a \emph{Least Recently Used} algorithm \cite{effelsberg1984principles}. As a result, it maintains a dictionary that is more suitable to the user, since it consists of more relevant recently typed words.

\noindent \textbf{Time-awareness:} \ours incorporates Time-Awareness (TA) in the PE, in order to incorporate user daily habits into the predictions. This is done by partitioning a day into  $T$ partitions, and by keeping a $WT$ per partition, represented as $WT_t$, $t \in [0,1,...,T]$. During a partition $t$, the respective $WT_t$ is active. Our experimental results (Section \ref{sec:exp}), showed that the invocation of the time feature improved the accuracy of \ours. 

\noindent \textbf{Multilingual Support:} PE is language agnostic, which means that in order to add a new language only the UI level has to be updated with the new character layout. This is a result of the way the PE handles the characters and the predictions, since $WT$s are independent of languages. Finally, in the case of a user typing in multiple languages, for each supported language there has to be a different instance of the PE.

\begin{algorithm}[h]
\caption{Time-Aware typing predictions}
\label{alg:treeGen}
\begin{footnotesize}
  {\bf Input:} \\
  ~~$ch$: Key typed. \\
  ~~$feedback$: Bad Prediction Feedback. \\
  \textbf{Return:} \\
  ~~$P$: List of predicted next keys along with their respective probability. \\
\end{footnotesize}
\begin{algorithmic}[1]
\begin{footnotesize}

  \STATE $time \leftarrow$ $clock$.$Now$

  \IF {$feedback$} 
    \STATE $idle \leftarrow \TRUE$
    \STATE $diffidence$()
  \ELSE
    \STATE $confidence$()    
  \ENDIF 

  \IF {$idle$}
    \STATE $cursor \leftarrow TimeTries$[$time$] 
    \STATE $cursor$.Add($ch$)
    \STATE $cursor \leftarrow cursor$[$ch$]  
    \IF {isWordSeparator($ch$)}
      \STATE $idle \leftarrow \FALSE$
      \STATE $TimeTries$[$time$] $\leftarrow TimeTries$[$time$].$Root$
      \RETURN $\emptyset$
    \ENDIF
  \ENDIF

  \STATE $cursor \leftarrow TimeTries$[$time$]
  \STATE $cursor \leftarrow cursor$[$ch$]
  \FOR {each $ch \in cursor$}
    \STATE $P \leftarrow P \cup$ popularity($ch$)
  \ENDFOR
  \IF {isWordSeparator($ch$)}
    \STATE $TimeTries$[$time$] $\leftarrow TimeTries$[$time$].$Root$
  \ENDIF

  \RETURN normalize($P$)

\end{footnotesize}
\end{algorithmic}
\end{algorithm}

\section{Evaluation}
\label{sec:exp}
\noindent \textbf{Dataset.} We utilized a dataset with messages typed through mobile devices. More specifically we collected Twitter messages and kept only the ones containing geo-location information. This information is only available when a user posts through GPS-enabled devices (smart phones, tablets, etc). We crawled Twitter using the Streaming API for the period between \textit{February 16, 2014} and \textit{April 6, 2014}. Tweets were collected using a bounding box around the United Kingdom, resulting in a dataset with over $27$ million tweets. We selected the top $100$ authors in terms of tweet count. In addition, we developed a simulator of the \ours engine (source code available here \cite{github:2013:sim}), in order to simulate the users' typing behavior using \ours. Retweets and hyperlinks were ignored, because they do not contain words and characters actually typed by the users. Instead, mentions to other users were taken into account, because they can be considered as nicknames that the user uses to refer to a friend. Finally, we manually filtered-out accounts that are automated systems (e.g. weather broadcasting accounts, bots, or news broadcasting accounts). Messages were sorted in a chronological order. We simulated users typing through \ours, by using multiple-sized training sets containing $0$ to $1500$ tweets, with a step of $50$ tweets. The test sets consist of the following $500$ messages. 

\noindent \textbf{Results:} We evaluated the predictive accuracy of \ours, by computing the hit ratio \cite{garay2006text} over how many predicted characters were returned in each prediction. The ideal prediction engine would always predict exactly one character which would be the correct one, resulting in a precision of $1$.  

For the experiments, we used the two following variations of \ours:
\begin{itemize}
\item \ours with Google's dictionary for the UK English language \cite{android:dic}, consisting of $160,722$ words, $Accelerator$-disabled, $TA$-disabled and $Pruning$-disabled.
\item \ours with no initial dictionary, $Accelerators$ disabled, $TA$-enabled and $Pruning$-enabled.
\end{itemize}

\begin{figure}[ht]
  \centering
  \subfigure[Precission]{\includegraphics[width=0.25\textwidth,natwidth=1166,natheight=875]{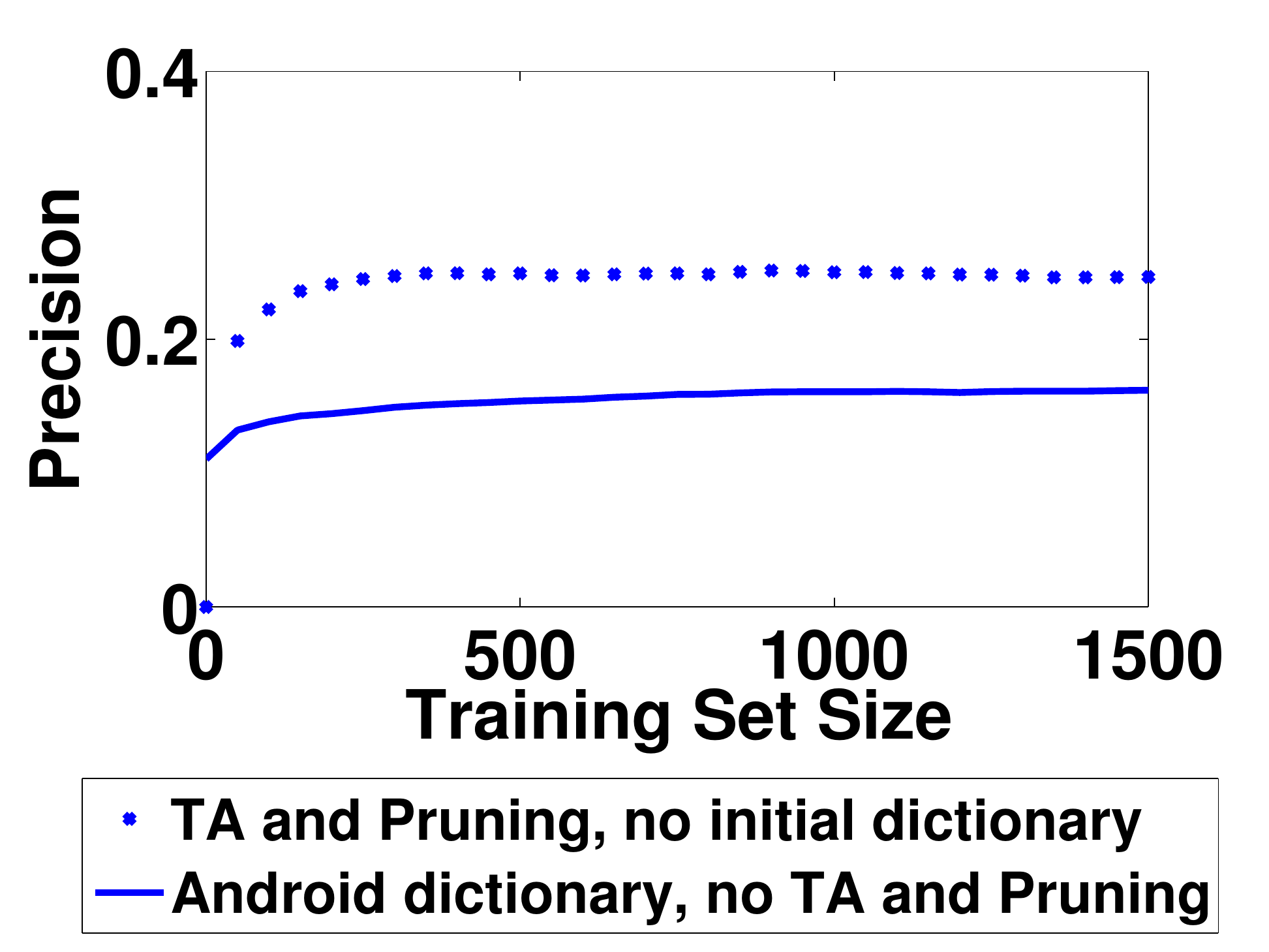}}  
  \centering
  \subfigure[Words learned]{\includegraphics[width=0.25\textwidth,natwidth=1166,natheight=875]{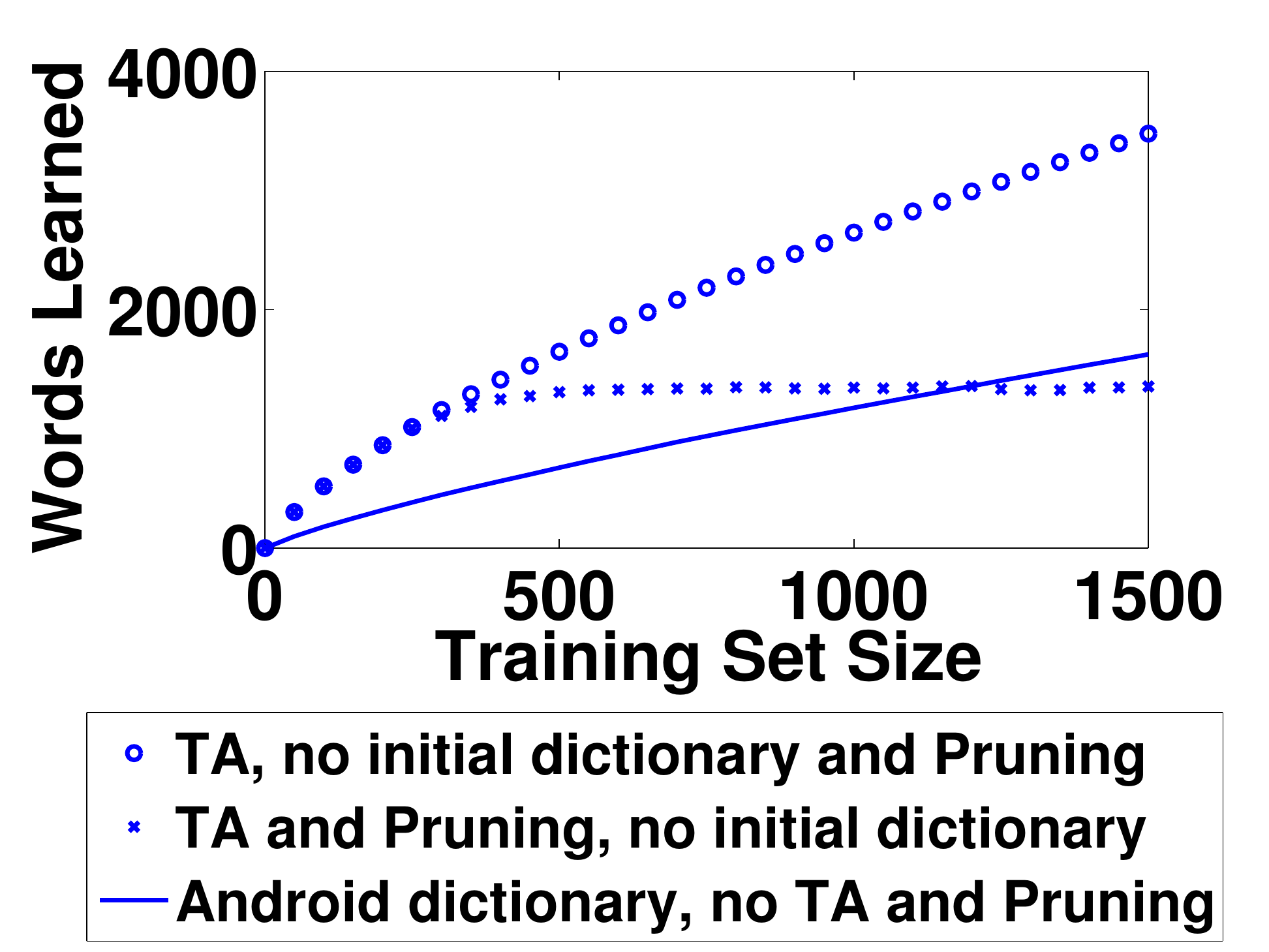}}
  \caption{Results} 
  \label{fig:pre_words}
\end{figure}

\begin{table}
\begin{tabular}{|c|c|c||c|c|c|}
\hline
\bfseries TA & \bfseries Pruning & \bfseries Prec & TA & \bfseries Pruning & \bfseries Prec \\\hline
\xmark & \xmark & .2387 & $T=14$ & \xmark & .2454 \\ \hline
$T=2$  & \xmark & .2426 & $T=15$ & \xmark & .2455 \\ \hline
$T=3$  & \xmark & .2418 & $T=16$ & \xmark & .2457 \\ \hline
$T=4$  & \xmark & .2426 & $T=17$ & \xmark & .2457 \\ \hline
$T=5$  & \xmark & .2432 & $T=18$ & \xmark & .2458 \\ \hline
$T=6$  & \xmark & .2435 & $T=19$ & \xmark & .2460 \\ \hline
$T=7$  & \xmark & .2439 & $T=20$ & \xmark & .2461 \\ \hline
$T=8$  & \xmark & .2442 & $T=21$ & \xmark & .2462 \\ \hline
$T=9$  & \xmark & .2446 & $T=22$ & \xmark & .2463 \\ \hline
$T=10$ & \xmark & .2448 & $T=23$ & \xmark & .2464 \\ \hline
$T=11$ & \xmark & .2450 & $T=24$ & \xmark & .2465 \\ \hline
$T=12$ & \xmark & .2452 & $T=24$ & \cmark & .2392 \\ \hline
$T=13$ & \xmark & .2454 & \xmark & \cmark & .2361 \\ \hline   
\end{tabular}
\caption{Mean Precision for \ours variations, without any $Accelerator$ and no initial dictionary}  
\label{tab:precision}
\end{table}

The dictionary approach, performed worse than \ours, which had no initial dictionary, even when using small training sets. Results are depicted in Figure \ref{fig:pre_words}a. The $TA$ approaches performed better than the simple ones and as the number of the time partitions increased, the precision also increased. The results of \ours with increasing number of time partitions can be found in Table \ref{tab:precision}. \ours without $Pruning$ performed worse than the version with $Pruning$, but the losses were covered from the gain of the $TA$.

Moreover, we studied the size of the words learned when $Pruning$ was disabled and we observed that the number of words was increasing with the training set size. Figure \ref{fig:pre_words}b depicts a comparison of the different methods with respect to the number of used words as a function of the training set size. Even for the Google's dictionary of 160,722 words, \ours continued learning new words as the training set increased, as depicted in Figure \ref{fig:pre_words}.

\section{Conclusion and Future Work}

In this paper we introduce \ours, an adaptive software keyboard for touchscreen devices that learns the user's dictionary and habits, in order to produce more accurate character predictions. During the demonstration, the users will be able to test \ours with a variety of touchscreen devices and settings by providing their Twitter username, so that \ours can accelerate learning of their personal dictionary. We plan to incorporate location features and knowledge exchanging between users that have similar behavior, e.g. users that are co-workers or are connected through a real-life relationship. Finally, we intend to use more $Acceleration$ sources, like chat messages, e-mails and social networking services.

An early prototype of \ours, without the \emph{Twitter Accelerator} and the $TA$ feature, is publicly available at Google's Play Store \cite{app:2013:wr}, with more than 9000 installations, a rating of 4 out of 5, among 72 users and 196 Google $+1$s. Source code is available at \cite{github:2011:wr}.

\section{Demonstration}

Attendants will be able to use two software keyboards:
\begin{enumerate}
	\item \textit{Time-Aware} \ours, after being $Accelerated$ by their Twitter account and 
	\item \ours without \textit{Time-Aware} or $Acceleration$ and with Google's Android dictionary \cite{android:dic} as initial dictionary
\end{enumerate}

and compare them by using them.
\bibliographystyle{abbrv}
\bibliography{anima}

\begin{thebibliography}{10}

\bibitem{apple:ios}
Apple.
\newblock ios 8.
\newblock \url{http://www.apple.com/ios/ios8/quicktype/}, Apr. 2014.

\bibitem{bi2013octopus}
X.~Bi, S.~Azenkot, K.~Partridge, and S.~Zhai.
\newblock Octopus: evaluating touchscreen keyboard correction and recognition
  algorithms via.
\newblock In {\em Proceedings of the SIGCHI Conference on Human Factors in
  Computing Systems}, pages 543--552. ACM, 2013.

\bibitem{effelsberg1984principles}
W.~Effelsberg and T.~Haerder.
\newblock Principles of database buffer management.
\newblock {\em ACM Transactions on Database Systems (TODS)}, 9(4):560--595,
  1984.

\bibitem{garay2006text}
N.~Garay-Vitoria and J.~Abascal.
\newblock Text prediction systems: a survey.
\newblock {\em Universal Access in the Information Society}, 4(3):188--203,
  2006.

\bibitem{android:dic}
Google.
\newblock Latinime.
\newblock
  \url{https://android.googlesource.com/platform/packages/inputmethods/LatinIME},
  June 2014.

\bibitem{gunawardana2010usability}
A.~Gunawardana, T.~Paek, and C.~Meek.
\newblock Usability guided key-target resizing for soft keyboards.
\newblock In {\em Proceedings of the 15th international conference on
  Intelligent user interfaces}, pages 111--118. ACM, 2010.

\bibitem{rodrigues2013improving}
E.~Rodrigues, M.~Carreira, and D.~Gon{\c{c}}alves.
\newblock Improving text entry performance on tablet devices.
\newblock {\em Intera{\c{c}}{\~a}o}, 2013.

\bibitem{github:2011:wr}
P.~Sakkos.
\newblock Writeright source code.
\newblock \url{https://github.com/PanosSakkos/write-right-app}, Dec. 2011.

\bibitem{github:2013:sim}
P.~Sakkos.
\newblock Prediction engine source code.
\newblock \url{https://github.com/PanosSakkos/anima}, Nov. 2013.

\bibitem{app:2013:wr}
P.~Sakkos.
\newblock Writeright keyboard (english).
\newblock
  \url{https://play.google.com/store/apps/details?id=panos.sakkos.softkeyboard.writeright},
  Sept. 2013.

\bibitem{damn:2014}
Userbase.
\newblock Damn you auto-correct!
\newblock \url{http://www.damnyouautocorrect.com}, June 2014.

\bibitem{willard1984new}
D.~E. Willard.
\newblock New trie data structures which support very fast search operations.
\newblock {\em Journal of Computer and System Sciences}, 28(3):379--394, 1984.

\bibitem{yin2013making}
Y.~Yin, T.~Y. Ouyang, K.~Partridge, and S.~Zhai.
\newblock Making touchscreen keyboards adaptive to keys, hand postures, and
  individuals: a hierarchical spatial backoff model approach.
\newblock In {\em Proceedings of the SIGCHI Conference on Human Factors in
  Computing Systems}, pages 2775--2784. ACM, 2013.

\end{thebibliography}

\end{document}